\documentclass[aps,prev,twocolumn,preprintnumbers,floatfix,nofootinbib]{revtex4-1}
\pdfoutput=1
\usepackage{amssymb,graphicx}
\usepackage{epstopdf}
\usepackage{mathrsfs}
\usepackage{amssymb}
\usepackage{verbatim}
\usepackage{color}
\usepackage{multirow}
\usepackage{amsmath}
\usepackage{subfig}
 \usepackage{slashed} 
\usepackage{float}
\usepackage[normalem]{ulem}
\usepackage{sidecap}
\usepackage{hyperref}
\usepackage{cancel}
\usepackage{enumerate}
\hypersetup{pdfstartview=FitV,colorlinks=true,linkcolor=blue,citecolor=red,filecolor=black,urlcolor=blue}

\usepackage[T1]{fontenc}
\usepackage[utf8]{inputenc}

\def\stacksymbols #1#2#3#4{\def\theguybelow{#2}
    \def\vp{\lower#3pt}
    \def\sp{\baselineskip0pt\lineskip#4pt}
    \mathrel{\mathpalette\intermediary#1}}

\def\intermediary#1#2{\vp\vbox{\sp
     \everycr={}\tabskip0pt
     \halign{$\mathsurround0pt#1\hfil##\hfil$\crcr#2\crcr
              \theguybelow\crcr}}}

\newcommand{\lslashslash}{%

  \raisebox{0.8ex}{%
    \scalebox{.7}{%
      \rotatebox[origin=c]{18}{$-$}%
    }%
  }%
}
\newcommand{\lslash}{%
  {%
   \vphantom{d}%
   \ooalign{\kern-.1em\smash{\lslashslash}\hidewidth\cr${\rm l}$\cr}%
   \kern.05em
  }%
}

\newcommand{\beq}{\begin{equation}}
\newcommand{\eeq}{\end{equation}}
\newcommand{\bea}{\begin{eqnarray}}
\newcommand{\eea}{\end{eqnarray}}

\newcommand{\gsim}{\lower.7ex\hbox{$\;\stackrel{\textstyle>}{\sim}\;$}}
\newcommand{\lsim}{\lower.7ex\hbox{$\;\stackrel{\textstyle<}{\sim}\;$}}

\newcommand{\mrm}{\mathrm}
\newcommand{\trh}{T_{\mathrm{RH}}}
\newcommand{\tmax}{T_{\mathrm{max}}}

\def\be{\begin{equation}}
\def\ee{\end{equation}}
\def\bea{\begin{eqnarray}}
\def\eea{\end{eqnarray}}

\def\m{\mu}
\def\n{\nu}

\def\sp{\;\;\;,\;\;\;}

\def\mrm{\mathrm}

\def\lsim{\raise0.3ex\hbox{$\;<$\kern-0.75em\raise-1.1ex\hbox{$\sim\;$}}}
\def\gsim{\raise0.3ex\hbox{$\;>$\kern-0.75em\raise-1.1ex\hbox{$\sim\;$}}}

\def\inbar{\,\vrule height1.5ex width.4pt depth0pt}

\def\IC{\relax\hbox{$\inbar\kern-.3em{\rm C}$}}
\def\IQ{\relax\hbox{$\inbar\kern-.3em{\rm Q}$}}
\def\IR{\relax{\rm I\kern-.18em R}}
 \font\cmss=cmss10 \font\cmsss=cmss10 at 7pt
\def\IZ{\relax\ifmmode\mathchoice
 {\hbox{\cmss Z\kern-.4em Z}}{\hbox{\cmss Z\kern-.4em Z}}
 {\lower.9pt\hbox{\cmsss Z\kern-.4em Z}}
 {\lower1.2pt\hbox{\cmsss Z\kern-.4em Z}}\else{\cmss Z\kern-.4em Z}\fi}

\def\tmax{T_{\rm max}}
\def\trh{T_{\rm RH}}

\def\comment#1{}
\def\to{\rightarrow}

\def\u1x{U(1)_X}
\newcommand{\nc}{\newcommand}
\nc{\LL}{L}
\nc{\vv}{\tilde{v}}
\nc{\ccdot}{\!\cdot\!}
\nc{\gsm}{G_{SM}}
\nc{\vfive}{\mathbf{5}\oplus\mathbf{\overline{5}}}
\nc{\vten}{\mathbf{10}\oplus\mathbf{\overline{10}}}
\nc{\zhol}{Z^{\rm hol}}
\nc{\xfb}{\,{\rm fb}}

\begin{document}

%\wideabs{
%\begin{flushright}
%
%
%\end{flushright}

\preprint{UMN--TH--4008/21}
\preprint{FTPI--MINN--21/01}

\vspace*{1mm}

\title{Gravitational production of dark matter during reheating}

\author{Yann Mambrini$^{a}$}
\email{yann.mambrini@th.u-psud.fr}
\author{Keith A. Olive$^{b}$}
\email{olive@physics.umn.edu}

\vspace{0.1cm}

\affiliation{
${}^a$ Universit\'e Paris-Saclay, CNRS/IN2P3, IJCLab, 91405 Orsay, France
 }
 \affiliation{
${}^b$ 
 William I.~Fine Theoretical Physics Institute, 
       School of Physics and Astronomy,
            University of Minnesota, Minneapolis, MN 55455, USA
}

\begin{abstract} 

We consider the direct $s$-channel gravitational production of dark matter
during the reheating process.  Independent of the identity of the dark matter candidate or its non-gravitational interactions, the gravitational process is always present and provides a minimal production mechanism.
During reheating, a thermal bath is quickly generated with a maximum temperature $\tmax$, and the temperature decreases as the inflaton continues to decay until the energy densities
of radiation and inflaton oscillations are equal, at $\trh$.  During these oscillations, $s$-channel gravitational production
of dark matter occurs. We show that the abundance of dark matter
(fermionic or scalar) depends primarily on the combination
$\tmax^4/\trh M_P^3$. We find that a sufficient density of dark matter can be produced over a wide range of dark matter masses: from a GeV to a ZeV. 

\end{abstract}

\maketitle

%%%%%%%%%

%%%%%%%%%%%%%%%%%%%%%%%%%%%%%%%%%%%%%%%%%%%%%%%%%%%%%%%%%%%%%%%%%%%%%%

\section{Introduction}

While we have considerable certainty in the existence of dark matter (DM),
its identity and interactions with the Standard Model are entirely unknown.
The lack of a signal in direct detection experiments \cite{XENON,LUX,PANDAX} sets
strong limits on the DM-proton cross section. Furthermore the lack of detection
at the LHC \cite{nosusy} 
also seems also to point towards more massive candidates and 
perhaps a more massive beyond-the-Standard Model sector than was originally envisioned \cite{hut,gunn,ehnos}. 

The mechanism by which dark matter particles populate the Universe is also unknown. Commonly, GeV-TeV
DM candidates are assumed to exist in equilibrium
as part of the thermal bath. As the temperature
falls below the DM mass, they drop out of equilibrium, 
and their relic density freezes out \cite{hut,swo,gg,gs,Arcadi:2017kky}. However, it is quite possible
that DM particles never attain thermal equilibrium.
They may either be too massive, or their interactions
with the Standard Model may be too weak. For example,
particles which interact with the Standard Model primarily through gravity, such as the gravitino
never achieve equilibrium, though they are produced
by the thermal bath at reheating after inflation \cite{nos,ehnos,kl,oss}. 
Very roughly, their abundance $Y \sim n_{3/2}/n_\gamma$
can be estimated from their production rate, $Y \sim \Gamma_p/H \sim \trh/M_P$ where $H$ is the Hubble parameter, $\trh$ is the reheating temperature after inflation, and $M_P=1/\sqrt{8\pi G_N}$ is the (reduced) Planck mass. 
This mechanism, now generally referred to as freeze-in, applies to a wider class of dark matter candidates known
as feebly interacting massive particles or FIMPs \cite{fimp,Bernal,Bernal:2017kxu,Bernal:2019mhf,Barman:2020plp}. 
Other examples include dark matter particles produced by 
the exchange of a massive $Z'$ \cite{Bhattacharyya:2018evo} or massive spin-2 \cite{Bernal:2018qlk}.

It is also possible that DM is produced in the decay of the inflaton, either directly \cite{egnop,grav2,Garcia:2017tuj,GKMO1} or radiatively \cite{Kaneta:2019zgw}. It has also been observed that
annihilation-like processes such as $\phi \phi \to S S$, where $\phi$ is the inflaton and $S$ is a dark matter scalar, mediated by gravity can produce a sufficient abundance of dark matter \cite{ema}. Indeed, the production of dark matter solely mediated by gravity
is a minimal contribution which is nearly model independent as we discuss in more detail below. 
The production of dark matter mediated by gravity from the thermal bath is subdominant \cite{Garny:2015sjg,Tang:2017hvq,Bernal:2018qlk,Chianese:2020yjo,Redi:2020ffc}. 

Often, reheating is characterized by a single temperature, $\trh$, which may be defined when
the energy density in the newly produced thermal bath becomes equal to the energy density still stored in inflaton oscillations. However, when one drops the approximation of instantaneous reheating, one finds that
initially, the Universe reheats to a potentially much higher temperature, $\tmax$, though very little of the total energy density of the Universe is in the form of radiation \cite{Giudice:2000ex,grav2,Garcia:2017tuj,Chen:2017kvz,GKMO1,Bernal:2020gzm}. 
For all models in which dark matter is produced during the reheating process after inflation, the dark matter abundance may be sensitive to the evolution
of the temperature between $\tmax$ and $\trh$. 
The purpose of this paper is to compute the minimal dark matter abundance produced solely through gravity during reheating taking into account the sensitivity of the result to details of the reheating process. 
 We will show that the bulk of the production of dark matter occurs at temperatures higher than the reheating temperature (when radiation begins to dominate the energy density) while the radiation bath begins to form. We stress that this process is always present, though other forms of production, including preheating, may also play a role and affect the final dark matter abundance.  Here, we will only consider the universal gravitational production channel.

The paper is organized as follows.
In the next section, we lay out the framework of our calculation. Starting with the universal coupling of gravity to the stress-energy tensor for either scalars or fermions, we compute the annihilation-like rate
for $\phi \phi \to S S$ (for scalar dark matter) and $\phi \phi \to \chi \chi$ in the case of fermionic dark matter. In section \ref{sec:dm}, we compute the dark matter abundance based on the detailed process of reheating between $\tmax$ and $\trh$. Then in section 
\ref{sec:uv}, we relate these results to possible
inflaton couplings to the Standard Model (which are responsible for reheating), and we draw our conclusions in \ref{sec:concl}.

\section{The framework}

\begin{figure}[ht]
\centering
\includegraphics[width=3.in]{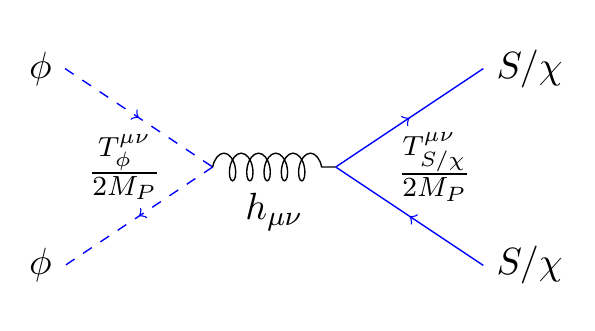}
\caption{\em \small Illustration of the production of dark matter through the gravitational scattering of the inflaton.}
\label{Fig:feynman}
\end{figure}

The universal interaction that surely exists between the inflaton and any dark sector is gravity. In particular,  the $s$-channel exchange of a graviton shown in Fig.~\ref{Fig:feynman} can be obtained from the Lagrangian (see e.g., \cite{hol}) 
\beq
{\cal L}=\frac{1}{2M_P}h_{\mu \nu}T^{\mu \nu}_\phi + \frac{1}{2M_P}h_{\mu \nu}T^{\mu \nu}_{S/\chi}
\eeq
where $h_{\mu \nu}$ is the metric perturbation corresponding to the graviton
and we consider either a scalar\footnote{We will consider a real scalar or Dirac fermionic as dark matter. Generalization to a complex scalar or vectorial dark matter is straightforward.}
$S$ or a fermion $\chi$ as dark matter, whose stress-energy tensors are given by 
\bea
T^{\mu \nu}_{X=\phi,S} &=&
\partial^\mu X \partial^\nu X-
g^{\mu \nu}
\left[
\frac{1}{2}\partial^\alpha X \partial_\alpha X-V(X)\right] \\
\label{Eq:tensors}
T^{\mu \nu}_\chi &=&
\frac{i}{4}
\left[
\bar \chi \gamma^\mu \overset{\leftrightarrow}{\partial^\nu} \chi
+\bar \chi \gamma^\nu \overset{\leftrightarrow}{\partial^\mu} \chi \right] \nonumber \\
&& -g^{\mu \nu}\left[\frac{i}{2}
\bar \chi \gamma^\alpha \overset{\leftrightarrow}{\partial_\alpha} \chi
-m_\chi \bar \chi \chi\right], 
\label{Eq:tensorf}
\eea
where $V(X)$ is the scalar potential for either the 
inflaton or scalar dark matter.

The amplitudes relevant for the computation of the processes $\phi(p_1)+\phi(p_2) \rightarrow \text{DM}^j (p_3)+\text{DM}^j(p_4)$ can be parametrized by
\begin{equation}
\mathcal{M}^{\phi j} \propto M_{\mu \nu}^\phi \Pi^{\mu \nu \rho \sigma} M_{\rho \sigma}^j \;, 
\end{equation}
where $j$ denotes the spin of the DM involved in the process and $j=0,1/2$. $\Pi^{\mu \nu \rho \sigma}$ denotes the propagator of the graviton with momentum $k = p_1+p_2$,
\begin{equation}
 \Pi^{\m\n\rho\sigma}(k) = \frac{\frac{1}{2}\eta^{\rho\n}\eta^{\sigma\m} + 
\frac{1}{2}\eta^{\rho\m}\eta^{\sigma\n} - \frac{1}{2}\eta^{\rho\sigma}\eta^{\m\n} }{k^2} \, .
\end{equation} 
The partial amplitudes, $M_{\mu \nu}^a$, can be expressed as
\bea 
M_{\mu \nu}^0 &=& \frac{1}{2}(p_{1\mu} p_{2\nu} + p_{1\nu} p_{2\mu} - g_{\mu \nu}p_1\cdot p_2 - g_{\mu \nu} m_{\phi}^2) \,, \nonumber\\ 
M_{\mu \nu}^{1/2} &=&  \frac{1}{4} {\bar v}(p_4) \left[ \gamma_\mu (p_3-p_4)_\nu + \gamma_\nu (p_3-p_4)_\mu \right] u(p_3) \;,
\eea
with a similar expression for scalar dark matter in terms of the dark matter momenta, $p_3, p_4$, and mass $m_S^2$.

It is then straightforward to compute the production rate of the dark matter.  Given the inflaton number density, $n_\phi = \rho_\phi/m_\phi$, the rate for the process depicted in Fig.~\ref{Fig:feynman} is
\bea
&&
\Gamma_{\phi \phi \rightarrow SS}
=\frac{\rho_\phi m_\phi}{1024 \pi M_P^4} \left( 1+\frac{m_s^2}{2 m_\phi^2} \right)^2 \sqrt{1-\frac{m_s^2}{m_\phi^2}}
\label{scalar}\\
&&
\Gamma_{\phi \phi \rightarrow \chi \chi}=\frac{\rho_\phi m_\chi^2}{4096\pi M_P^4 m_\phi}
\left(1-\frac{m_\chi^2}{m_\phi^2}\right)^{3/2}
\label{fermion}
\eea
Note the difference in behavior in the expressions for fermionic and scalar dark matter, especially in the mass dependence. On dimensional grounds, both are proportional
to $n_\phi m_\phi^2/M_P^4 = \rho_\phi m_\phi/M_P^4$.
However, our rates in Eqs.~(\ref{scalar}) and (\ref{fermion}) correspond to s-wave scattering within the condensate. As a result, in the case of a Dirac fermion-antifermion pair in the final state, we require a spin flip leading to a suppression by a factor $(m_\chi/m_\phi)^2$, making the rate proportional to $\rho_\phi m_\chi^2/ m_\phi M_P^4$. A similar expression for the rate producing scalars was found in \cite{ema}.

\section{Dark matter production}
\label{sec:dm}

In many models of inflation, after the period of exponential expansion, the inflaton begins a series of oscillations about a minimum. During the initial stages of oscillations, the Universe expands as if it were dominated by non-relativistic matter. Inflaton decays
begin the process of reheating \cite{dl,nos}. While we assume that decay products thermalize rapidly \cite{rapid}, we do not assume that the decay is instantaneous. Instead, we include the effects due to the evolution of the temperature of the radiation bath
from its initial creation to a temperature $\tmax$ until it begins to dominate the expansion at $\trh$.

Let us start by considering the evolution of the radiation bath using the conservation equation for radiation (produced by the decay of the inflaton),
\beq
\frac{d\rho_R}{dt} + 4 H \rho_R = \Gamma_\phi \rho_\phi \, ,
\eeq
where $\Gamma_\phi$ is the inflaton decay rate, and $\rho_R$ and $\rho_\phi$ are the energy densities of radiation and the inflaton respectively. 
This equation can be rewritten as
\beq
\frac{dX}{da} = \frac{\Gamma_\phi \rho_\phi}{H} a^3 \, ,
\label{Xa}
\eeq
where $X = \rho_R a^4$, $a$ is the cosmological scale factor and $H \equiv {\dot a}/a \approx \sqrt{\rho_\phi}/\sqrt{3} M_P$ up to the epoch of radiation domination. Assuming inflaton oscillations about a quadratic potential\footnote{For similar results about a more general potential see \cite{GKMO2}.},  $\rho_\phi = \rho_e (a_e/a)^3$,  where $\rho_e$ is the energy density stored by the inflaton at the end of the inflationary phase, and Eq.~(\ref{Xa}) is easily integrated to give
\beq
\rho_R = \frac{2\sqrt{3}}{5} \Gamma_\phi M_P \sqrt{\rho_e} \left( \frac{a_e}{a} \right)^{3/2} \left[1-\left( \frac{a_e}{a} \right)^{5/2} \right] = \alpha T^4 ,
\label{rhoR}
\eeq
where $\alpha = g_T \pi^2/30$ and $g_T$ is the number of relativistic degrees of freedom, at temperature $T$.
Thus, in the case of non-instantaneous reheating, when 
$a \gg a_e$, and as long as the inflaton dominates the energy density, the temperature of the radiation bath falls as $T \propto a^{-3/8}$ as energy is injected into the bath from inflaton decay \cite{Giudice:2000ex,Garcia:2017tuj,Kaneta:2019zgw}.
For $a \gg a_e$, with the help of Eq.(\ref{rhoR}) we can then write 
\beq
H(T) \simeq \frac{\sqrt{\rho_\phi}}{\sqrt{3}M_P} =
\frac56 \frac{\alpha}{\Gamma_\phi M_P^2} T^4 \, .
\eeq
Within the same limit we can parametrize the reheating temperature by
\beq
\trh = \left( \frac{4}{3 \alpha c^2} \right)^{1/4} \left( \Gamma_\phi M_P\right)^{1/2}
\eeq
where $c = 1$ if we define the reheat temperature by $t_{\rm RH}^{-1} = \frac32 H = \Gamma_\phi$ and $c=5/3$ if we define $\trh$ by $\rho_\phi (\trh) = \rho_R (\trh)$. Here, we have chosen the latter
which can easily be derived by rewriting Eq.(\ref{rhoR}) in the limit $a\gg a_e$
\beq
\rho_R\simeq\frac{2 \sqrt{3}}{5}\Gamma_\phi M_P \sqrt{\rho_\phi(\trh)}  
\left(\frac{a_{\rm RH}}{a}\right)^{3/2}\, ,
\eeq
so that $\rho_{\phi}(\trh)=\rho_R(\trh)$ implies that
\beq
\rho_R(\trh) = \alpha \trh^4 \simeq\frac{12}{25}\Gamma_\phi^2 M_P^2.
\eeq
Then
the expression for $H(T)$ in a Universe dominated by the inflaton is given by \cite{Kaneta:2019zgw,GKMO2} 
\beq
H(T)=\sqrt{\frac{\alpha}{3}}\frac{T^4}{\trh^2 M_P} \, ,
\eeq
 and 
\beq
\rho_\phi= \rho_R(\trh) \left(
\frac{a_{\rm RH}}{a} \right)^{3} =
\alpha \frac{T^8}{\trh^4}  \, .
\label{rhophiofT}
\eeq

When the limit $a \gg a_e$ cannot be applied, the
reheating temperature is the solution 
of the equation
\beq
\sqrt{\rho_{\rm RH}} = \alpha^{1/2} \trh^2 =
\frac{2 \sqrt{3}}{5}\Gamma_\phi M_P \left[1-\left(\frac{\rho_{\rm RH}}{\rho_e}\right)^{5/6}\right] \, ,
\label{Eq:rhorexact}
\eeq
where $\rho_{\rm RH}=\rho_\phi(T_{RH})=\rho_R(\trh)$.

Note that the maximum temperature attained, $T_{\rm max}$, can be found from the extremum of Eq.~(\ref{rhoR}) with respect to $a$.
We obtain
\beq
\frac{a_{\rm max}}{a_e}=\left(\frac{8}{3}\right)^{2/5}
\eeq
implying
\beq
\rho_R(a_{\rm max})=\frac{\sqrt{3}}{4} \Gamma_\phi M_P\sqrt{\rho_e}\left(\frac38 \right)^{3/5} = \alpha \tmax^4 \, .
\label{Eq:rhomax}
\eeq
Combining Eqs.~(\ref{Eq:rhorexact}) and (\ref{Eq:rhomax}) gives
\beq
\left(\frac{\tmax}{\trh}\right)^4=  \frac{25 \sqrt{\rho_e}}{16 \sqrt{3} \Gamma_\phi M_P} 
\left( \frac{3}{8}\right)^{3/5} \left[1-\left( \frac{a_e}{a_{\rm RH}} \right)^{5/2} \right]^{-2} \, .
\label{Eq:ratioy0}
\eeq

The relic abundance of dark matter is obtained by solving the Boltzmann equation for the number density of dark matter particles
\beq
\frac{dn_j}{dt}+3Hn_j=R_j(T) \, .
\label{boltz}
\eeq

Writing $Y_j = n_j a^3$, 
Eq.~(\ref{boltz}) can be simplified to
\beq
\frac{dY_j}{da}= \frac{R_j(T) a^2}{H}
\label{Eq:y}
\eeq
where $R_j(T)=\frac{\rho_\phi}{m_\phi}\Gamma_{\phi \phi \rightarrow jj }$ is the production rate (per unit volume and unit time) and the yield, $Y_j$ is proportional to the number of dark matter quanta produced in the comoving frame.  
It is important to note that because $R_j \propto \rho_\phi^2 \propto T^{16}$, there is a very strong temperature dependence in Eq.~(\ref{Eq:y}), making the result sensitive to the maximum temperature attained.

Using Eq.~(\ref{rhophiofT}), it is straightforward to 
integrate Eq.~(\ref{Eq:y}) between $a_e$ and $a_{\rm RH}$ 
We find
\beq
Y_j(\trh) = \frac{2 \gamma_j  M_P}{\sqrt{3}m_\phi} \rho_e^{3/2} a_e^3 \left[ 1-\left(\frac{a_e}{a_{\rm RH}} \right)^{3/2} \right] \,,
\label{Ysol0}
\eeq
where $\gamma_j = \Gamma_{\phi\phi\to jj}/\rho_\phi$.
The yield can also be written in terms of $\rho_R$,
and the number density becomes
\beq
n_j(\trh) = \frac{2 \gamma_j M_P}{\sqrt{3}m_\phi}  \alpha^{3/2} \trh^6  \left[ \left( \frac{a_{\rm RH}}{a_e} \right)^{3/2}-1 \right] \,.
\label{Ysol1}
\eeq
Noting that we can write $(a_e/a_{\rm RH}) = (a_e/a_{\rm max})(a_{\rm max}/a_{\rm RH}) = (3/8)^{2/5} (\tmax/\trh)^{8/3}$, the density can be expressed as
\beq
n_j(\trh) = \frac{2 \gamma_j M_P}{\sqrt{3}m_\phi}  \alpha^{3/2} \trh^2  \left[ \left( \frac83 \right)^{3/5} \tmax^4 - \trh^4 \right] \,.
\label{Ysol2}
\eeq

Equation ~(\ref{Ysol2}) is valid so long as $a_{\rm RH} > a_{\rm max}$. For sufficiently large $\Gamma_\phi$, the radiation energy density will equal the inflaton oscillation energy density when $T= \tmax$, and thus $\trh = \tmax$.
At larger $\Gamma_\phi$, we can write
\beq
n_j(\trh) = \frac{2 \gamma_j M_P}{\sqrt{3}m_\phi}  \alpha^{3/2} \trh^6  \left[ \left( \frac{\sqrt{\rho_e}}{\alpha^{1/2} \trh^2} \right)-1 \right] \,,
\label{Ysol3}
\eeq
as the temperature $\tmax$ is never attained.

The fraction of critical density in dark matter can be written as
\beq
\Omega_j h^2 = 1.6 \times 10^8 \left( \frac{g_0}{g_{\rm RH}}\right)\left( \frac{n_j(\trh)}{\trh^3}\right)
\left( \frac{m_j}{1~\mrm{GeV}} \right) \, ,
\label{Eq:omegah2}
\eeq
where the numerical factor is $\pi^2 n_\gamma(T_0) /2\zeta(3)\rho_c$, $g_0 = 43/11$, $g_{\rm RH} = 427/4$, and $n_j(\trh)$ is obtained from either  Eq.~(\ref{Ysol2}) or Eq.~(\ref{Ysol3}) for $a_{\rm RH} > a_{\rm max}$ or $a_{\rm RH} < a_{\rm max}$, respectively. 
We finally obtain
\bea
\frac{\Omega_sh^2}{0.1} & \simeq &
\left( \frac{\trh}{10^{10}}\right)^3
\left(\frac{\tmax/\trh}{100}\right)^4
\left(\frac{m_S}{1.75\times 10^{10}}\right) \nonumber \\
&
\times & \left( 1+\frac{m_s^2}{2 m_\phi^2} \right)^2 \sqrt{1-\frac{m_s^2}{m_\phi^2}} ;\; a_{\rm RH} > a_{\rm max} \, , \label{Eq:omegas} \\
\frac{\Omega_sh^2}{0.1} & \simeq &
\left( \frac{\trh}{10^{10}}\right)
\sqrt{\frac{\rho_e}{10^{60}}}
\left(\frac{m_S}{1.9\times 10^{9}}\right) \nonumber \\
&
\times & \left( 1+\frac{m_s^2}{2 m_\phi^2} \right)^2 \sqrt{1-\frac{m_s^2}{m_\phi^2}};\; a_{\rm RH} < a_{\rm max} \, ,
\label{Eq:omegas1}
\eea
for scalar dark matter and
\bea
\frac{\Omega_\chi h^2}{0.1} & \simeq &
\left(\frac{\trh}{10^{10}}\right)^3
\left(\frac{\tmax/\trh}{100}\right)^4
\left(\frac{m_\chi}{4.0\times 10^{12}}\right)^3
\nonumber
\\
&
\times &
\left(\frac{3\times 10^{13}}{m_\phi}\right)^2
\left(1-\frac{m_\chi^2}{m_\phi^2}\right)^{3/2} ;\; a_{\rm RH} > a_{\rm max} \, , \label{Eq:omegachi} \\
\frac{\Omega_\chi h^2}{0.1} & \simeq &
\left(\frac{\trh}{10^{10}}\right)
\sqrt{\frac{\rho_e}{10^{60}}}
\left(\frac{m_\chi}{1.9\times 10^{12}}\right)^3
\nonumber
\\
&
\times &
\left(\frac{3\times 10^{13}}{m_\phi}\right)^2
\left(1-\frac{m_\chi^2}{m_\phi^2}\right)^{3/2} ;\; a_{\rm RH} < a_{\rm max} \, ,
\label{Eq:omegachi1}
\eea
for fermionic dark matter.
Units of energy are given in GeV (and GeV$^4$ for $\rho_e$).  Note that these expressions are approximations which are valid so long as $a_{\rm RH}$ is not very close to either $a_{\rm max}$ or $a_e$. More generally, the relic density is obtained from Eq.~(\ref{Eq:omegah2})
with either Eq.~(\ref{Ysol2}) or Eq.~(\ref{Ysol3}). As expected, there is a stronger dependence of the relic abundance on the mass of the dark matter in the fermionic case due to its production rate.

\begin{figure}[ht]
\centering
\includegraphics[width=3in]{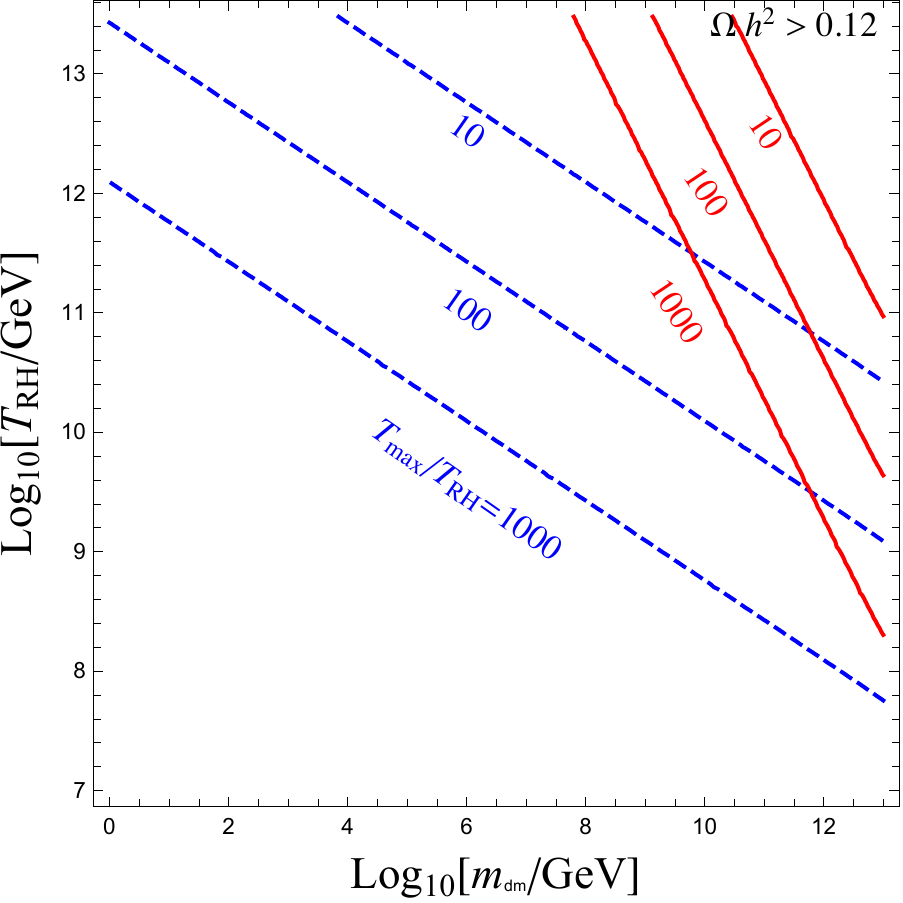} \\
\caption{\em \small Points yielding the Planck relic density $\Omega_{S,\chi} h^2=0.12$ for scalar (blue, dashed) and fermionic (red, solid) dark matter, in the  
($m_{S,\chi}$, $\trh$) plane for several values of ${\tmax}/{\trh}$ as labeled. }
\label{Fig:omegaSS}
\end{figure}

We show in Fig.~\ref{Fig:omegaSS} the allowed parameter space in the ($m_{S,\chi}$, $T_{\rm RH}$) plane for scalar (blue, dashed) and fermionic (red, solid) dark matter for several values of $\tmax/\trh$. The lines correspond to
values of $\Omega h^2 = 0.12$ \cite{Planck} for each choice of $\tmax/\trh$. In each case, $a_{\rm RH} > a_{\rm max}$.
All points above the lines are excluded because they lead to an overabundance. We see, for example, that GeV-ZeV dark matter can be obtained with reasonable values of $\trh$ and $\tmax$ by pure gravitational production through inflaton scattering.
We also see that due to the additional mass suppression $(m_\chi/m_\phi)^2$, for fermionic DM, it is necessary to consider
higher fermionic masses ($\gtrsim 100$ PeV for $\tmax/\trh < 1000$) to achieve $\Omega_\chi h^2 = 0.12$.  Note that we have assumed $m_\phi = 3 \times 10^{13}$ GeV in Fig.~\ref{Fig:omegaSS} and the kinematic suppression when $m_s \ge m_\phi$ would appear to the right of the region plotted.

\section{Coupling to the Standard Model}
\label{sec:uv}

In the previous section, we concentrated on results for a given reheating and maximum temperature reached by the thermal bath during the reheating process. In any concrete model of inflation,
these temperatures can be traced to two quantities: the coupling of the inflaton to the Standard Model, which ultimately determines the reheating temperature, and the initial energy density of the inflaton at the end of the inflationary phase, which allows one to determine the amount of energy available for the thermal bath, and hence the maximum temperature. If one supposes a simple effective coupling of the inflaton to the Standard Model fermions of the type
\beq
{\cal L}_{\phi-SM}^y=y~\phi \overline f f,
\label{Eq:lagrangianf}
\eeq
leading to 
\beq
\Gamma_\phi = \frac{y^2}{8\pi} m_\phi \, , 
\eeq
and using Eq.~(\ref{Eq:ratioy0}) we can write the ratio 
\beq
\left(\frac{\tmax}{\trh}\right)^4=  \frac{25 \pi \sqrt{\rho_e}}{2 \sqrt{3}y^2m_\phi M_P} 
\left( \frac{3}{8}\right)^{3/5} \left[1-\left( \frac{a_e}{a_{\rm RH}} \right)^{5/2} \right]^{-2} .
\label{Eq:ratioy}
\eeq
  In models which are dominated by a quadratic term after inflation, we expect $\rho_e \sim m_\phi^2 M_P^2$ and for example, in the Starobinsky model of inflation \cite{staro}, $\rho_e^{1/4} \sim 0.65 m_\phi^{1/2} M_P^{1/2} = 5.5 \times 10^{15}$ GeV \cite{egno5}. For sufficiently large coupling, $y \simeq .67 (\rho_e^{1/4}/10^{15} {\rm GeV})$, $\tmax/\trh = 1$, and reheating is nearly instantaneous.
 If reheating is primarily due to a coupling of the inflaton to bosons, ${\cal L}_{\phi-SM}^\mu=\mu~\phi bb$, equivalent results are obtained by substituting $y \to \mu/m_\phi$. 

Similarly, we can also compute $\trh$ by combining Eqs.~(\ref{Eq:rhorexact}) and ~(\ref{Eq:lagrangianf}):
\beq
\trh^4=\left(\frac{3y^4}{400 \alpha \pi^2}\right) m_\phi^2 M_P^2 \left[ 1 - \left( \frac{\alpha \trh^4}{\rho_e} \right)^{5/6} \right]^2  \, .
\label{Eq:trhy}
\eeq
For sufficiently large $\rho_e$, the second term in the bracket can be neglected, otherwise, $\trh$ can be solved for iteratively. 
We can now express the relic density in terms of the coupling $y$ using Eqs.~(\ref{Eq:ratioy}) and (\ref{Eq:trhy}) in Eqs.~(\ref{Ysol2}) and (\ref{Ysol3}).
When $a_{\rm RH} \gg a_{\rm max}$, we can write relatively compact expressions by
combining Eqs.(\ref{Eq:omegas}), (\ref{Eq:omegachi}), (\ref{Eq:ratioy}) and (\ref{Eq:trhy}) and one obtains
for $m_{S,\chi} \ll m_\phi$
\bea
\frac{\Omega_S h^2}{0.1}&=&\frac{y}{10^{-7}}
\sqrt{\frac{\rho_e}{10^{64}}}
\sqrt{\frac{3\times 10^{13}}{m_\phi}}
\frac{m_S}{5.2\times 10^9}
\label{oh2sy}
\\
\frac{\Omega_\chi h^2}{0.1} & = & \frac{y}{10^{-7}}
\sqrt{\frac{\rho_e}{10^{64}}}
\left(\frac{m_\chi}{2.6 \times 10^{12}}\right)^3
\left(\frac{3\times 10^{13}}{m_\phi}\right)^{3/2} \, ,
\eea
where as before dimensions for mass are in GeV, with GeV$^4$ for $\rho_e$. 
We see that for a given Standard Model coupling ($y = 10^{-7}$ for example), the dark matter mass needed to reach a reasonable relic abundance is
much higher in the fermionic case than in the scalar case,  bearing in mind the minimality of the model being considered. 

\begin{figure}[!ht]
\centering
\includegraphics[width=3in]{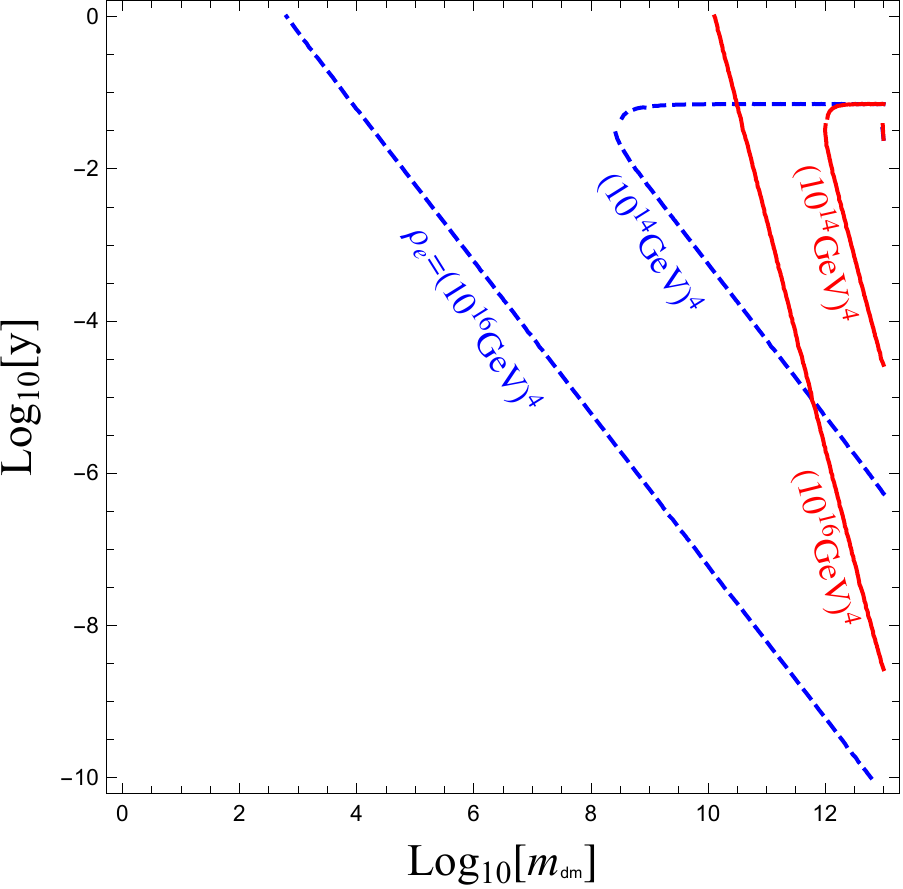}
\caption{\em \small Points in the 
($m_{S,\chi}$, $y$) plane yielding the Planck-determined relic density $\Omega_{S,\chi} h^2=0.12$ in the case of a scalar dark matter (blue,dashed) and a fermionic dark matter (red, solid) for two values of $\rho_e$ as labeled.}
\label{Fig:omegaSS2}
\end{figure}

We plot in Fig.~\ref{Fig:omegaSS2} the required relation between $m_{S,\chi}$ and $y$ to obtain $\Omega_{S,\chi} h^2 = 0.12$.

Values of $\Omega h^2 > 0.12$  are obtained for points to the right of the negatively sloped lines and below the horizontal lines for $\rho_e^{1/4} = 10^{14}$ GeV.
Note that for $\rho_e^{1/4} = 10^{14}$ GeV, the
curves begin to bend when the second term in the bracket in Eq.~(\ref{Eq:ratioy})
approaches 1. For small values of $y$, $\tmax \gg \trh$, and
the relic density in Eq.~(\ref{Eq:omegas}) scales as $T_{\rm RH}^3 (T_{\rm max}/T_{\rm RH})^4 m_s$. Since 
$(T_{\rm RH})^3 \sim y^3$, $(T_{\rm max}/T_{\rm RH})^4 \sim y^{-2}$, we see that
$y \sim m_S^{-1}$ for fixed relic density.
At larger values of $y$, when $a_{\rm RH} < a_{\rm max}$, the density in Eq.~(\ref{Eq:omegas1}) scales as $\trh m_s$. However, we can see from Eq.~(\ref{Eq:trhy}), that 
$\trh$ reaches its maximal value when $\alpha \trh^4 \lesssim \rho_e$. At these values of $y$ (which are not much larger than $.67 (\rho_e^{1/4}/10^{15}~{\rm GeV})$ - the value needed to make $\trh = \tmax$), the number density obtained from Eq.~(\ref{Ysol3}) begins to decrease. Thus a fixed energy density which is proportional to $n_j m_j$, requires a rapidly larger value of the dark matter mass. This accounts for the curve bending to the horizontal at large values of $m_j$.
Physically, at these values of $y$, the rate of inflaton decay exceeds the rate of dark matter production from scattering, and production through the gravitational process ceases. For $\rho_e^{1/4} > 1.5 \times 10^{15}$ GeV, this behavior occurs only at $y > 1$. 
Once again, the region with kinematic suppression lies to the right of the region shown in the figure. 
 
It is also possible that inflaton scattering into Standard Model scalars affects the reheating process, and the value of $\tmax$ in particular \cite{GKMO2}. For example, a Lagrangian contribution of 
\beq
{\cal L}_{\phi-SM}^{\sigma}=\sigma \phi^2b^2
\label{Eq:lagrangianb}
\eeq
where $b$ represents a Standard Model boson. In this case, one still requires the Yukawa coupling (\ref{Eq:lagrangianf}) to complete the reheating process (so that the energy density in radiation comes to dominate over the energy density in $\phi$) \cite{GKMO2}. The interaction generated by (\ref{Eq:lagrangianb}) can
alter the maximum temperature $\tmax$ for sufficiently large values of $\sigma$. In this case, 
\beq
\left(\frac{\tmax}{\trh}\right)^4=\frac{6400\pi \sigma^2 \rho_e^{3/2}}{\sqrt{3}y^4m_\phi^5 M_P} \left(\frac23 \right)^{18} 
\label{tmaxtrhs}
\eeq
giving
\bea
\frac{\Omega_S h^2}{0.1} & = & \frac{10^{-6}}{y}
\left(\frac{\sigma}{10^{-9}}\right)^2
\left(\frac{\rho_e}{(10^{16})^4}\right)^{3/2}
\left(\frac{3 \times 10^{13}}{m_\phi}\right)^{7/2} \nonumber \\
&& \times \left(\frac{m_S}{1.2\times 10^5} \right)
\eea
and
\bea
\frac{\Omega_\chi h^2}{0.1} & = &
\frac{10^{-6}}{y}\left(\frac{\sigma}{10^{-9}}\right)^2
\left(\frac{\rho_e}{(10^{16})^4}\right)^{3/2}
\left(\frac{m_\chi}{7.6\times 10^{10}}\right)^3
\nonumber
\\
&&
\times
\left(\frac{3 \times 10^{13}}{m_\phi}\right)^{11/2}
\eea
Note that for a given value of $\sigma$, the relic density increases with decreasing $y$. This is valid only so long as $\tmax > \trh$. The limiting value of $y$
is found by setting Eq.~(\ref{tmaxtrhs}) equal to one.

\begin{figure}[!ht]
\centering
\vskip .2in
\includegraphics[width=3.in]{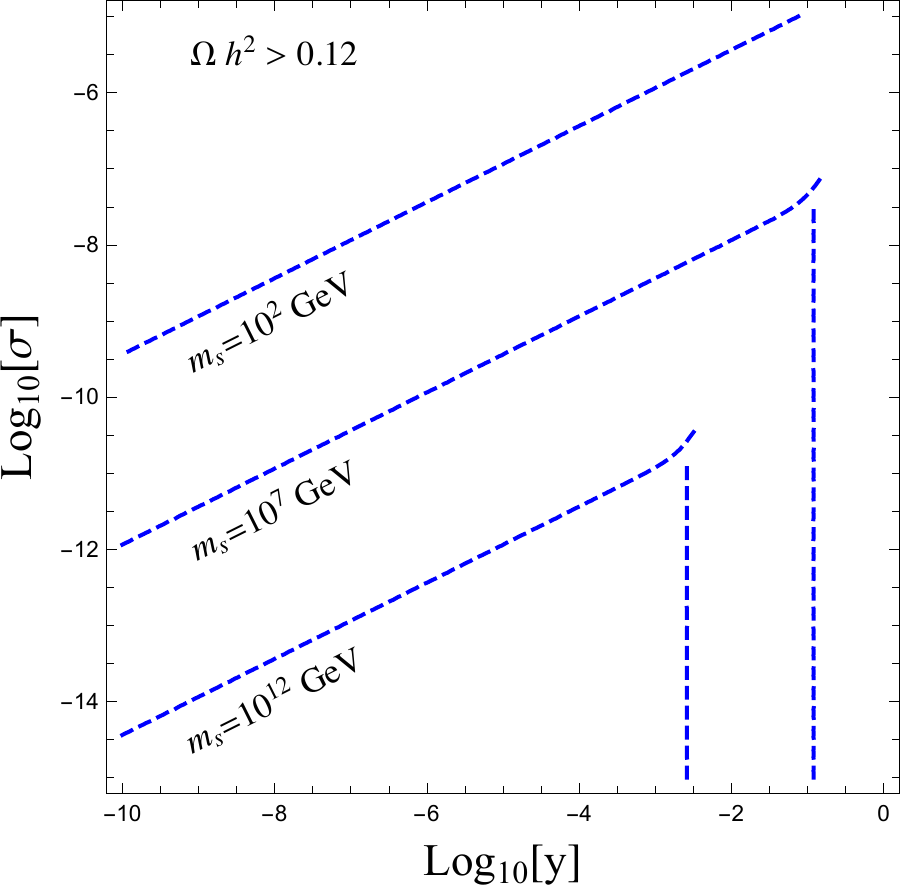}
\caption{\em \small Points respecting Planck constraint $\Omega_S h^2=0.12$ in the case of scalar dark matter, in the plane 
($y$, $\sigma$) for different values of $m_S$.}
\label{Fig:omegaSSFbis}
\end{figure}

We show in Fig.(\ref{Fig:omegaSSFbis}) the parameter space allowed in the plane ($y$, $\sigma$) in the case of scalar dark matter for different values of its mass $m_\chi$.
Note that it is possible to have quite low values of $y$ (and thus $\trh$) while still being able to produce dark matter in sufficient amounts due to the value of $\sigma$ generating a high maximum temperature $\tmax$, and thus a large production rate. 
The maximum value for $y$ occurs when $\tmax = \trh$ in Eq.~(\ref{tmaxtrhs}) where the relic abundance depends only on $\trh$ (and is thus independent of $\sigma$) as we can see in the plot. The same curves for fermionic dark matter are obtained for $m_\chi = 1.5\times 10^9 (m_S/{\rm GeV})^{1/3}$.

\section{Conclusions}
\label{sec:concl}

We have derived the conditions for producing sufficient dark matter
from inflaton scattering during reheating by $s$-channel graviton
exchange. This process is always present independent of the model of inflation.  We have shown that the final abundance of dark matter
depends not only on the reheating temperature, but also
on the maximum temperature and hence on the detailed evolution of the reheating process. During the exit from exponential expansion, many models of inflation begin a period of oscillations leading to reheating. At the onset of these oscillations, the inflaton density is high and the leading contribution to dark matter production occurs at the start of reheating at $\tmax$.  This represents an absolute minimal amount of dark matter production and it contributes independent of any interactions the dark matter may have with the Standard Model
(or another dark sector if present).

\noindent {\bf Acknowledgements. }  This work was made possible by Institut Pascal at Universit\'e
Paris-Saclay with the support of the P2I research departments and 
the P2IO Laboratory of Excellence (program ``Investissements d'avenir''
ANR-11-IDEX-0003-01 Paris-Saclay and ANR-10-LABX-0038). This project has received support from the European Union’s Horizon 2020 research and innovation programme under the Marie Sk$\lslash$odowska-Curie grant agreement No 860881-HIDDeN and the CNRS PICS MicroDark. The work of K.A.O.~was supported in part by DOE grant DE-SC0011842  at the University of
Minnesota.

\end{document}